\documentstyle[twocolumn,aps]{revtex}
\begin{document}
\draft
\title{Vortex phase diagram for mesoscopic superconducting disks} 
\author
{V.A. Schweigert \cite {*:gnu}, F.M. Peeters \cite {f:gnu}, and P.S. Deo}
\address{\it  Departement Natuurkunde, Universiteit Antwerpen (UIA),\\
Universiteitsplein 1, B-2610 Antwerpen, Belgium}
\date{\today}
\maketitle
\begin{abstract}
Solving numerically the 3D non linear Ginzburg-Landau (GL) equations,
we study equilibrium and nonequilibrium phase transitions
between different superconducting states
of mesoscopic disks which are thinner than the coherence length
and the penetration depth.
We have found a smooth transition from a multi--vortex superconducting 
state to a giant vortex state
with increasing both the disk thickness and the magnetic field. A 
vortex phase diagram is obtained which 
shows, as function of the magnetic field,
a re-entrant behavior  between
the multi--vortex and the giant vortex state.

\end{abstract}

Recently,  mesoscopic superconductivity  has attracted
much attention in  view of phase transitions in confined systems
with sizes comparable to the coherence ($\xi$) and penetration ($\lambda$)
lengths. While the type of bulk superconductors
is only  determined by the value of
the Ginzburg-Landau parameter $\kappa=\lambda/\xi$,
the experimental observations \cite{geim} and
the numerical simulations \cite{deo,schweigert} of magnetization
of mesoscopic thin disks have shown that
the {\it type} and the {\it order} of those transitions
between different superconducting states and between
the superconducting and the normal state  depends crucially on
the disk radius $R$ and the thickness $d$. With increasing the disk radius
the second-order
reversible phase transition
observed for small disk radii are replaced by first-order transitions
with a jump in the magnetization. In previous theoretical investigations
of superconductivity in such mesoscopic disks \cite{deo,schweigert}
only the giant vortex states with  fixed total
angular momentum $L$ were considered with an axially symmetric 
order parameter. It is well known \cite{gennes} that for type-II
superconductors ($\kappa>1/\sqrt{2}$), the triangular Abrikosov
vortex lattice is energetically favorable in the range $H_{c1}<H<H_{c2}$.
Since the effective London penetration depth $\Lambda=\lambda^2/d$
increases considerably in thin disks with $d\ll \lambda$ one would expect
the appearance of the Abrikosov multi--vortex state even in disks
made from a material with $\kappa<1/\sqrt{2}$, like e.g. the $Al$ disks
studied in Refs.\cite{deo,schweigert}. By analogy with
classical particles confined by an external potential \cite{bedanov},
the structure of a finite number of vortices
should differ from a simple triangular arrangement
and allow for different metastable states.
Using the London approximation Fetter \cite{fetter} calculated
the critical field $H_{c1}$ for flux penetration into a disk.
For a superconducting cylinder the multi--vortex clusters, containing
up to four vortices, were simulated by Bobel \cite{bobel}.
Using the  method of images and the London approximation,
Buzdin and Brison \cite{buzdin} have considered
vortex structures in small $R\ll\Lambda$ disks and found a 
classical particle ringlike arrangement \cite{bedanov} of vortices.
In the present Letter we study the transition from the giant vortex
state to this multi--vortex configuration for thin superconducting
disks within the nonlinear Ginzburg--Landau (GL) theory.

 We consider a superconducting disk immersed in an insulator media
with a perpendicular uniform magnetic field $H_0$. 
For thin disks ($d\ll \xi,\lambda$) we found \cite{deo,schweigert}
that it is allowed to average
the GL equations over the disk thickness.
Using dimensionless variables and the London gauge $div \vec A=0$ for
the vector potential $\vec A$,
we write the system of GL equations in the following form
\begin{equation}
\label{eq1}
\left(-i\vec \nabla_{2D} -\vec A\right)^2\Psi=
\Psi (1-|\Psi|^2),
\end{equation}
\begin{equation}
\label{eq2}
-\triangle_{3D} \vec A=\frac{d}{\kappa^2}\delta(z) \vec j_{2D},
\end{equation}
\begin{equation}
\label{eq3}
\vec j_{2D}=\frac{1}{2i}
\left(\Psi^*\vec \nabla_{2D} \Psi-\Psi \vec\nabla_{2D}\Psi^*\right)
-|\Psi |^2\vec A,
\end{equation}
with the boundary condition 
$(-i\vec \nabla_{2D}-\vec A)\Psi|_{r=R}=0.$
Here the distance is measured in units of the coherence length
$\xi$, the vector potential in $c\hbar/2e\xi$, and the magnetic field in
$H_{c2}=c\hbar/2e\xi ^2=\kappa \sqrt{2}H_c$. The disk is placed in
the plane ($x,y$), the external magnetic field is directed along
the $z$-axis,
the indices $2D$, $3D$
refer to two-dimensional and three-dimensional operators, and
$\vec j_{2D}$ is the density of superconducting current. To solve the
system of Eqs.(\ref{eq1}-\ref{eq2}) we apply a finite-difference
representation of the order parameter and the vector potential on a
uniform cartesian space grid $(x,y)$ and use the link variable approach
\cite{kato}.
In our simulations we took typically
a grid spacing of $0.15\xi$.
To find the steady-state solution of  the GL equations
we add to the  LHS  of  Eqs. (\ref{eq1})  and (\ref{eq2})  the  time 
derivatives of
the order parameter and the vector potential, respectively, and use
an iteration procedure based on the Gauss-Seidel technique to find $\Psi$.
The vector potential is obtained with the fast Fourier transform technique.
For this purpose we set the condition for the vector potential 
$\vec A_{|x|=R_s,|y|=R_s}=H_0(x,-y)/2$
at the boundary of a larger space grid $(R_s=4R)$, where the vector
potential created by the superconducting currents is much less than the 
external vector potential.

The giant vortex state is characterized by the total angular momentum
$L$ through $\Psi=\psi(\rho)exp(iL\phi)$,
where $\rho$, $\phi$ are the cylindrical coordinates.
An arbitrary superconducting state
is generally a mixture of different angular harmonics. Nevertheless,
we can introduce an analog to the total angular momentum which
is still a good quantum number. 
Choosing circular loops at the
periphery of the disk we find that the effective angular momentum
$L=\Delta \phi/2\pi$ does not depend on the loop radii $\rho_l$  when
it is in some range $\rho_l=(0.8\div1)R$. This allows us to characterize unambiguously the different
superconducting states. Note, that the effective angular
momentum is in fact nothing else then the number of vortices
in the disk.

To find the different vortex configurations, which include 
the metastable states, we search for the steady-state solutions of
Eqs.(\ref{eq1}-\ref{eq2})
starting from different initial conditions which were generated randomly. 
Once we obtained these configurations for some value of the magnetic field we
increase/decrease slowly the magnetic field as long as the number of
vortices remains unchanged. Comparing the dimensionless Gibbs free energies
$F=V^{-1}\int (2(\vec A-\vec A_0)\vec j_{2d}-|\Psi|^4)d\vec r$, where
integration is performed over the disk volume $V$, and $\vec A_0$ is
the vector potential of the external uniform magnetic field, we found
the ground state. If the system has sufficient time to equilibrate,
the system will be in the ground state and we obtain
phase transitions between
different ground states. If the latter condition 
is not satisfied the system can remain in a metastable
state until this state disappears or becomes unstable with respect to small
perturbations in the order parameter and/or the magnetic field.
Such nonequilibrium phase transitions lead to
hysteresis in the magnetization. The barriers separating the metastable
and the ground state are discussed in Refs.\cite{bean,zhang}.
The system remains in the metastable state if the barrier height is
much larger than the temparature. Instead of finding the barrier
heights, which is a more cumbersome problem, we search for the critical
magnetic fields corresponding to the disappearance
of such barriers.
Here, we started from the Meissner (normal) state
and increased (decreased) slowly the magnetic field 
(typically with steps of $0.01H_{c2}$) from the
zero (nucleation) magnetic field. 

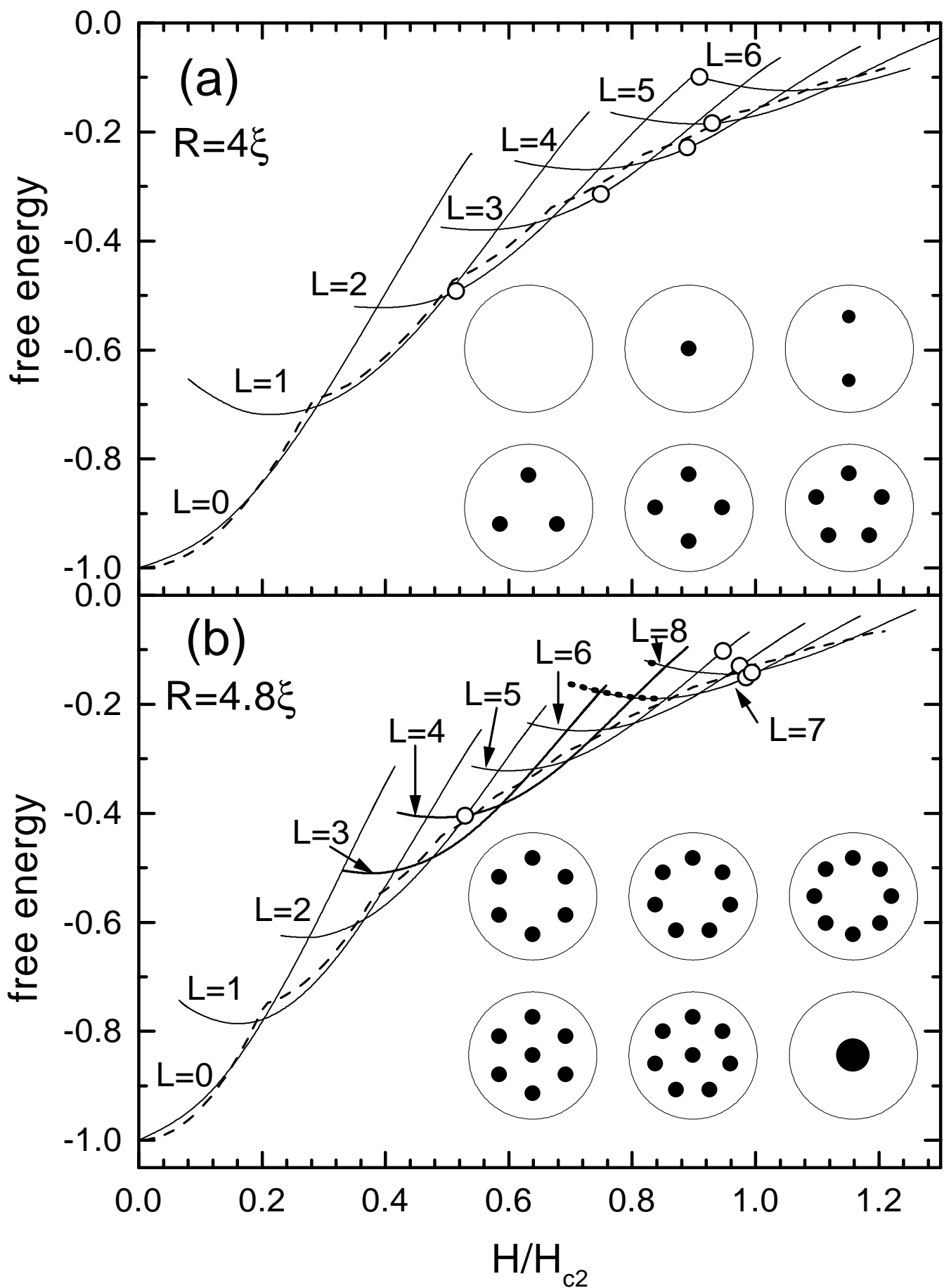
\begin{figure}
\caption{
The free energy of configurations with different number of vortices
$L$ for two disk radii $R=4\xi$ (a) and $R=4.8\xi$ (b) in the case of zero
disk thickness and $\kappa=0.28$.
The open circles indicate the transition from a multi--vortex
to  a giant vortex state (right side of circle). 
The dotted curves correspond to the free energy
of metastable configurations with a vortex inside a poligon for $L=7,8$~(b).
The dashed curves are the results of our approximate analytical
calculations. The insets show the possible vortex configurations.}
\label{fig1}
\end{figure}

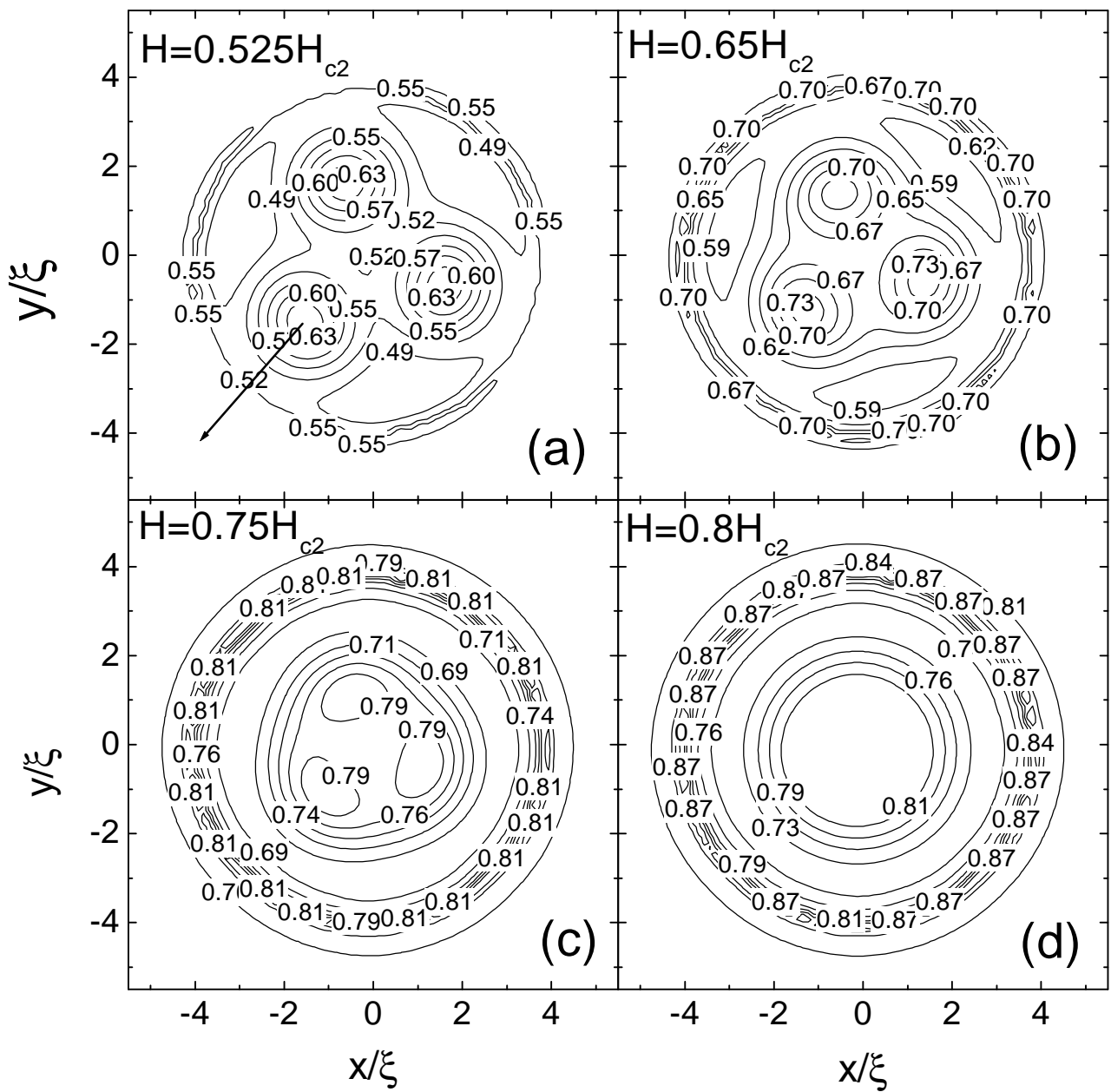
\begin{figure}
\caption{
Contourplot of the magnetic field distribution in the disk plane ($z=0$)
for the case of a three vortex state and 
for different applied magnetic fields $H_0=0.525H_{c2}$~(a),
$H_0=0.65H_{c2}$~(b), $H_0=0.75H_{c2}$~(c), and $H_0=0.8H_{c2}$~(d).
We took $R=4\xi$ and $d=0.5\xi\kappa^2$.
For the lowest magnetic field (a) the
three--vortex configuration is metastable and a small
decrease in the magnetic field leads to a transition to the two vortex state
due to the expulsion of a vortex through the disk boundary indicated by the
arrow in (a).} 
\label{fig2}
\end{figure}

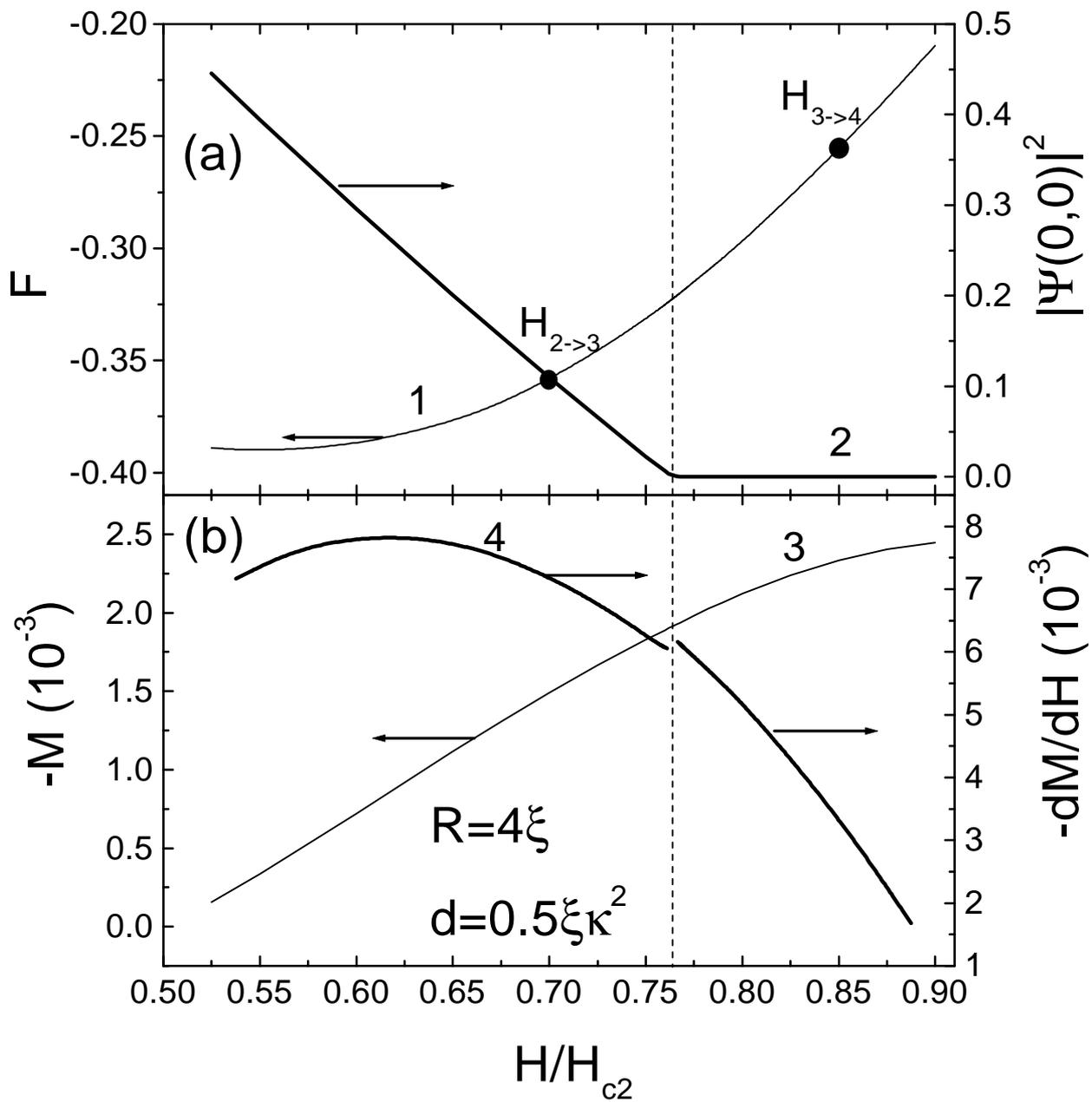
\begin{figure}
\caption{
The free energy (1), the square of the order parameter in the disk
center (2), the magnetization (3), and its first derivative (4)
for a disk with $R=4\xi$, $d=0.5\xi\kappa^2$ which is in the vortex state
with $L=3$. The dashed vertical line shows the transition from the
multi--vortex state to the giant vortex state. In (a) the solid circles
indicate  
the points at which the equilibrium phase transitions $2\rightarrow 3$ and
$3\rightarrow 4$ occur.}
\label{fig3}
\end{figure}

The free energies of the different vortex configurations are shown in
Fig.~\ref{fig1} for  zero disk thickness and for two disk radii:
 (a) $R=4\xi$, and (b) $R=4.8\xi$.
 For $R=4\xi$, the vortex configuration can consist
up to six vortices which are arranged on the edge of an ideal polygon.
 The pentagon and hexagon
vortex clusters are always metastable states for $R=4\xi$.
With increasing  disk radius the
allowed number of vortices  increases  leading  to the  appearance  of 
polygons
with a vortex inside it when $L>7$. We found that these structures
have a larger energy 
as compared to the ideal polygons. This result
differs from the London approximation of Ref.\cite{buzdin} where
the vortices were treated
 as an ensemble of
interacting classical point pseudoparticles in which
 the closed packed structures
are more preferable even for $L\ge 6$.
The present work clearly shows the limited validity of the
approach of Ref.\cite{buzdin} for the case of superconducting disks.
Another
unexpected feature is observed when the magnetic field is further increased.
While the model of classical particles predicts a decrease in
the intervortex distance followed by the appearance of a new vortex,
our simulations show, as a rule,
a gradual transition from a {\it multi--vortex state} to a {\it giant vortex
state} (Fig.~\ref{fig2}). In principle, the latter may be a metastable state.
The positions of these
transitions are indicated by circles in Fig.~\ref{fig1}. 
For a larger disk radius $R=4.8\xi$,
we also observe transitions between different
multi--vortex states.
Performing a Fourier analysis of the order parameter
we find that a multi--vortex state corresponding to an ideal polygon
presents a mixture of harmonics
$\Psi\approx \psi_0(\rho)+\sum_k\psi_k(\rho)exp(ikL\phi)$ with a rather small
contribution of higher ($k>2$) harmonics. This allows us to find
approximately the free energy of different vortex configurations in
thin disks $R\ll\Lambda$ in which we can neglect the distortion of the
magnetic field. For this purpose we take the order parameter as a
superposition of only two states
$\Psi\approx C_0^{1/2}\zeta_0(\rho)+C_L^{1/2}\zeta_L(\rho)exp(iL\phi)$,
where $\zeta_0$, $\zeta_L$ are the eigenfunction of the linearised first GL
equation (\ref{eq1}) for different angular momenta
\cite{schweigert,moschchalkov}. Substituting this expansion into
Eq.(\ref{eq1})
we obtain the following set of non-linear equations for the
coefficients $C_0$ and $C_L$
\begin{eqnarray}
\label{model}
\lambda_0C_0=a_{11}C_0^2+a_{12}C_0C_L,
\\ \nonumber
\lambda_LC_L=a_{12}C_0C_L+a_{22}C_L^2,
\end{eqnarray}
with $a_{11}=<\zeta_0^2|\zeta_0^2>$, $a_{12}=2<\zeta_0^2|\zeta_L^2>$,
$a_{22}=<\zeta_L^2|\zeta_L^2>$,
where $\lambda_0$, $\lambda_L$ are the eigenvalues of the
linearised GL equation \cite{schweigert},
and $<f_0|f_1>$ refers to the matrix element $V^{-1}\int f_0f_1d\vec r$.
Besides the two trivial solutions
$C_L=0$, $C_0=\lambda_0/a_{11}$ and
$C_0=0$, $C_L=\lambda_L/a_{22}$ which correspond to
the Meissner state and the giant vortex state, respectively, it is
possible to have another solution with
$C_0=(\lambda_0a_{22}-\lambda_La_{12})/D$,
$C_L=(\lambda_La_{11}-\lambda_0a_{12})/D$, $D=a_{11}a_{22}-a_{12}^2$
corresponding to a multi--vortex state.
 The free
energy of the ground state obtained using this approach,
is shown in Fig.~\ref{fig1} by the dashed curve, which is in good
agreement with the results of our simulations (full curves).
In order to discuss the multi--vortex $\leftrightarrow$ giant vortex transition,
we anylized the stability of
the obtained solutions with respect to small perturbations of the coefficients
$C_0$, $C_L$ (see Ref.\cite{schweigert}).
The stability conditions 
for the multi--vortex state and the giant vortex state are
$\lambda_L<\lambda_{\star}=\lambda_0a_{22}/a_{12}$ and
$\lambda_L>\lambda_{\star}$, respectively. Consequently for fixed $L$,
there is an unique solution with a reversible
transition from the multi--vortex state to the
giant vortex, which occurs with increasing
magnetic field when $\lambda_L=\lambda_{\star}$. At this critical
point the multi--vortex state coincides with the giant vortex state and
consequently there is no jump in the magnetization. But
the derivatives of the coefficients $C_0$, $C_L$ 
are discontinuous and correspondly we expect a discontinuty in
the first derivative of the magnetization. This is also confirmed by
our numerical simulations where in Fig.~\ref{fig3} the region around the $L=3$
multi--vortex to giant vortex transition is shown.

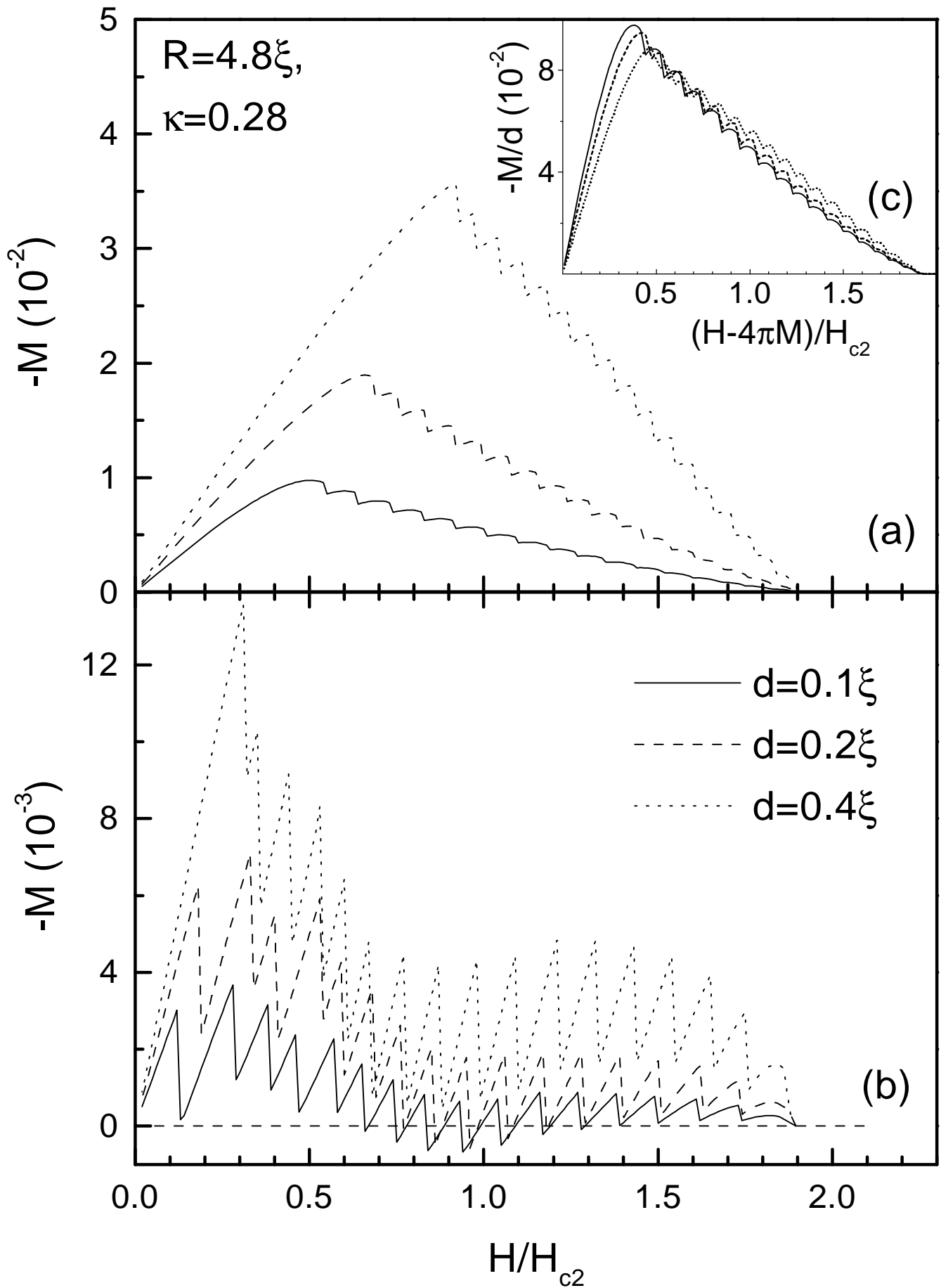
\begin{figure}
\caption{
Magnetization of a disk in increasing (a,c) and decreasing
(b) magnetic fields for $R=4.8\xi$, $\kappa=0.28$
and different disk radii.}
\label{fig4}
\end{figure}

The magnetization of the disk $M=\int (H-H_0)d\vec r/4\pi VH_{c2}$
with $R=4.8\xi$ is shown in Fig.~\ref{fig4} for
increasing (a) and decreasing (b) magnetic field.
Jumps in the magnetization correspond to transitions between states
with  different number of vortices at the magnetic fields where the states 
cease to exist (see Fig.~1). With  
increasiing disk thickness the demagnetization effect increases and 
as a result the transition points shift to larger magnetic fields, but the
number of jumps remains the same. 
The average magnetic field in the disk can be estimated
as $\langle H\rangle\approx H-4\pi M$ and the relative magnetizations $M/d$ as
a function of $\langle H\rangle$ is `almost' an universal curve for all disk thicknesses
(Fig.~\ref{fig4}(c)). For increasing magnetic field and disk thicknesses
$d=0.2,~0.4\xi$ all transitions occur between giant vortex states.
For the thinner disk $d=0.1\xi$ we observe the following sequence of transitions
$0\rightarrow 1\rightarrow 2_g\rightarrow 3_m\rightarrow 3_g
\rightarrow 4_m\rightarrow 4_g\rightarrow 5_g\rightarrow 6_g...$,
where the lower index ($g,m$) corresponds to giant and multi--vortex
states, respectively. For decreasing magnetic field 
multi--vortex states appear with a larger number of vortices. The sequence of
transitions between multi--vortex states start from $L=7$ and $L=8$ for
$d=0.2\xi$ and $d=0.1\xi$, respectively. For the thicker disk ($d=0.4\xi$),
a multi--vortex state appears only for $L=3,2$ just before the expulsion of
a vortex from the disk. As is evident from Fig.~\ref{fig4}(b), the 
appearance of positive magnetization observed experimentally 
can also be explained within the 3D GL approach without the consideration of
pinning effects. But the lowest energy state, i.e. the equilibrium state, has
always a negative magnetization.

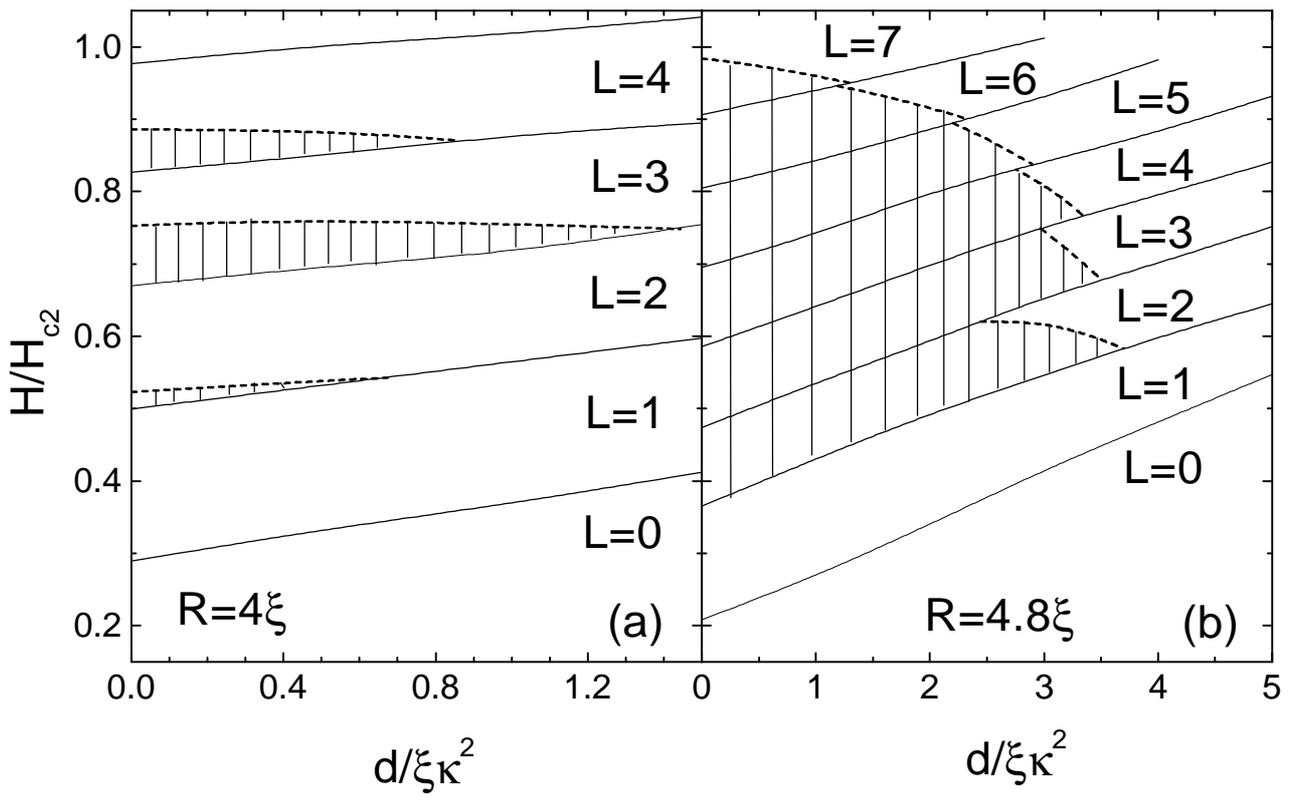
\begin{figure}
\caption{
The vortex phase diagram for two different disk radii $R=4\xi$~(a)
and $R=4.8\xi$ (b). The shaded area corresponds to the multi--vortex state}
\label{fig5}
\end{figure}

To distinguish quantitatively the giant vortex state 
with the multi--vortex states with the
same number of vortices we consider the value of the order
parameter $|\Psi|^2$ 
in the center of the disk which is nothing else than the density of
Cooper pairs. We find
that this parameter, which is zero for a giant vortex state,
goes almost linearly to zero  when the
magnetic field approaches some critical value (see 
Fig.~\ref{fig3} (a), the thick curve).
Therefore, the magnetic field
obtained by linearly interpolating $|\Psi(0,0)|^2$ to zero
defines the transition
from a multi--vortex state to a giant vortex state.
As mentioned above, our simulations show that this transition is reversible
and exhibits a discontinuous derivative in the magnetization.
Having the free energies of different vortex configurations we are
able to construct  
an equilibrium vortex phase diagram which is shown in Fig.~\ref{fig5} for two disk radii
$R=4\xi$ and $R=4.8\xi$, respectively. The solid curves separate the
regions with  different number of vortices and the dashed curves show
the boundaries between the multi--vortex and the giant vortex states.
For $L=1$ the single vortex state and the giant vortex state are identical. 
The shaded regions correspond to the multi--vortex states. The
superconducting to normal transition occurs for $H/H_{c2}\approx 1.9$
which is outside the plotted region. Notice that the multi--vortex area
in the phase diagram reduces in size with incresing disk thickness
and it disappears in the limit of thick disks where only the giant
vortex state survives. Thus for type I superconductors the
multi--vortex state is favored with decreasing disk thickness and
increasing disk radius. This behaviour can be understood as follows: 
with increasing radius the energy difference
between different $L$-states decreases and consequently it becomes
possible to build a lower energy multi-vortex state out of a linear
combination of giant vortex states. For decreasing radius this is
more difficult to do and there exists a critical radius below which no
multi--vortex states have the lowest energy.
To observe the multi--vortex $\leftrightarrow$ giant vortex transition one
should investigate the derivative of the magnetization or the vortex
configuration itself which can be done by using e.g. a magnetic force
microscope. 

This work is supported by the Flemish Science Foundation (FWO-VL)
through project 5.0277.97, the project INTAS-93-1495-ext and the
``Interuniversity Poles of Attraction Program - Belgian State, Prime Minister's
Office - Federal Office for Scientific, Technical and Cultural Affairs''.
One of us (FMP) is a research director with the FWO-VL.

\newpage

\end{document}